\documentclass[aps,a4paper,preprint]{revtex4}%
\usepackage{amsfonts}
\usepackage{amsmath}
\usepackage{amssymb}
\usepackage{graphicx}
\usepackage{setspace}%
\providecommand{\U}[1]{\protect\rule{.1in}{.1in}}
\begin{document}
\title{Crack propagation under static and dynamic boundary conditions}
\author{Yuko Aoyanagi and Ko Okumura}
\affiliation{Physics Department, Faculty of Science, Ochanomizu University}
\date{\today}

\begin{abstract}
Velocity jumps observed for crack propagation under a static boundary
condition have been used as a controlling factor in developing tough rubbers.
However, the static test requires many samples to detect the velocity jump. On
the contrary, crack propagation performed under a dynamic boundary condition
is timesaving and cost-effective in that it requires only a single sample to
monitor the jump. In addition, recent experiments show that velocity jump
occurs only in the dynamic test for certain materials, for which the velocity
jump is hidden in the static test because of the effect of stress relaxation.
Although the dynamic test is promising because of these advantages, the
interrelation between the dynamic test and the more established static test
has not been explored in the literature. Here, by using two simulation models,
we elucidate this interrelation and clarify a universal condition for
obtaining the same results from the two tests, which will be useful for
designing the dynamic test.

\end{abstract}
\maketitle

\section{Introduction}

Crack propagation is a crucial factor for controlling the toughness of
materials. For crack-propagation in elastomers, a remarkable phenomenon,
called velocity jump \cite{Tsunoda2000}, has recently been revisited and has
attracted considerable attention, which include experimental
\cite{MoridhitaUrayama2016PRE,morishita2017crack}, numerical
\cite{kubo2017velocity}, and analytical
\cite{sakumichi2017exactly,Okumura2018} studies. The velocity jump refers to a
sharp jump in the velocity of crack propagation, typically from 0.1 mm/s to 1
m/s, as a function of the energy release rate $G$, which is an increasing
function of the applied displacement $\varepsilon$. (In the linear case, $G$
is proportional to $\varepsilon^{2}$). The experiment is conventionally
performed under a static boundary condition (fixed-grip condition): crack
propagation starts at an initial displacement, while the displacement is kept
fixed during the crack propagation.

The analytical study \cite{sakumichi2017exactly} as well as the numerical
study \cite{kubo2017velocity} suggest that the physical origin of the velocity
jump is the glass transition at the crack tip \cite{KuboSakumichi}. The ratio
of the velocities before and after the jump is only four orders of magnitude
at most, but the region in which glass transition occurs is very localized
near the tip, causing a significant reduction in the characteristic length
scale; these two effects are combined to attain a significant change (of
nearly nine orders of magnitude) in strain rate at the crack tip required for
glass transition \cite{Okumura2018}.

Recently, it is reported that the velocity jump is not observed for a
semi-crystalline polymer \cite{Takei2018} as a result of performing the
crack-propagation test under the static boundary condition. However, more
recently, the velocity jump is successfully observed for the same
semi-crystalline polymer when the crack-propagation test is performed under a
dynamic boundary condition \cite{TomizawaJump2}.

In the dynamic test, the change in the velocity of crack propagation is
monitored when the sheet sample is extended at a constant speed in the
direction perpendicular to the direction of crack propagation. Because of this
dynamic boundary condition, the stress relaxation is minimized in the dynamic
test (in the static test, we generally have a preparation time for giving a
fixed strain before starting crack propagation), which leads to the
observation of the velocity jump. In other words, the dynamic test is more
sensitive for detecting the velocity jump for certain polymers and thus
applicable to a wider range of materials than the static test.

In addition to this advantage concerning the sensitivity, the dynamic test is
timesaving and cost-effective. This is because the dynamic test requires one
sample (to be broken) in order to obtain a single point in the
velocity-displacement plot. On the contrary, one complete
velocity-displacement curve is obtained from a single sample in the dynamic test.

As seen above, the dynamic test has strong advantages over the static test for
detecting the velocity jump. However, to date, there have been no studies
which discuss the relation between the results obtained from the static and
dynamic tests. Here, we elucidate this relation through a numerical study. We
use two simple viscoelastic models appropriate for examining crack propagation
\cite{aoyanagi2017,aoyanagiNonlinear}. One model is a spring-bead model based
on Voigt model. Note that analysis of such a simple model is always important
to know the basic properties. The other model is another spring-bead model, in
which viscous dissipation is introduced by a friction force proportional to
the bead velocity. We show in the Appendix rheological response of the model
to show that the model can appropriately describe essential features of
polymer rheology. From the results obtained from these two models, we clarify
simple and universal conditions under which the two tests provide equivalent
information. The results shall be useful for designing the dynamic test as a
clever substitute for the static test.

\section{Simulation models}

\begin{figure}[h]
\includegraphics[width=0.5\textwidth]{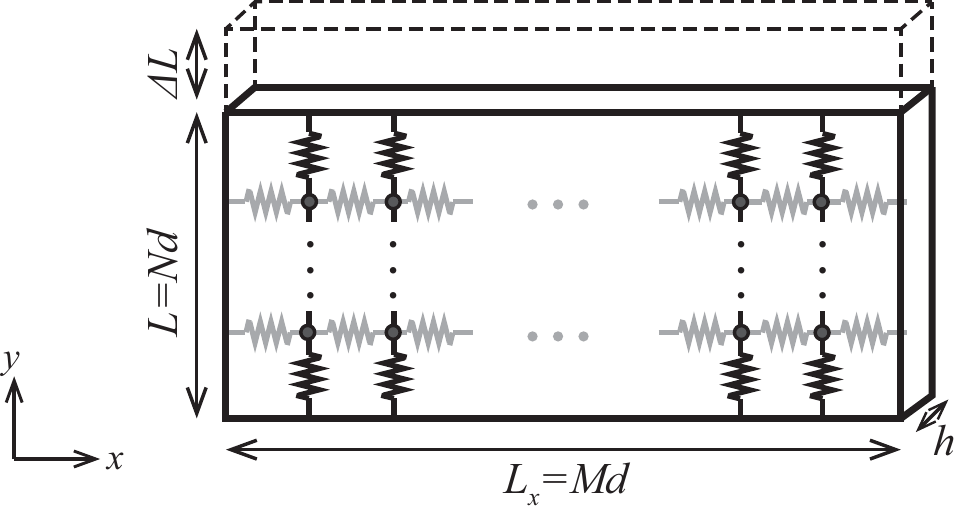}\caption{Schematic illustration
showing the arrangement of springs and beads in a network structure. $N$
springs are arranged in series in the $y$ direction and $M$ springs in the $x$
direction. Beads are located on nodal points of the springs.}%
\label{Fig1}%
\end{figure}

In order to represent the network structure in polymer materials, we consider
a simple two dimensional square-lattice network, as shown in Fig. \ref{Fig1}.
Beads are arranged on nodal points on a square lattice with lattice spacing
$d$; each bead is connected to the four nearest neighbors by springs of spring
constant $k$. The bead positions initially located at the lattice point
$\mathbf{m}=(i,j)$ are labeled by $\mathbf{x}_{\mathbf{m}}$. Let us define the
vector $\Delta\mathbf{x}_{\mathbf{mm}^{\prime}}$ as the extension vector from
the natural length of spring connecting $\mathbf{m}$ and $\mathbf{m}^{\prime}$
(see Sec. \ref{s0} of Appendix for the details, where we introduce the index
$s$ ($s=1,2,3,4$), for which $s=1$ and 3 correspond to the shear force and
$s=2$ and 4 to the tensile force in the $y$ direction); $\mathbf{m}^{\prime}$
is the index of one of the nearest neighbor sites of the site represented by
$\mathbf{m}$.

In the present study, we consider the following two types of equation of
motion:%
\begin{align}
\gamma\frac{d}{dt}\Delta\mathbf{x}_{\mathbf{m}}+k\Delta\mathbf{x}%
_{\mathbf{m}}  &  =\mathbf{F}_{\mathbf{m}}\label{single0}\\
\gamma\frac{d}{dt}\mathbf{x}_{\mathbf{m}}+k\Delta\mathbf{x}_{\mathbf{m}}  &
=\mathbf{F}_{\mathbf{m}}\label{multi0}\\
\Delta\mathbf{x}_{\mathbf{m}}  &  =\sum_{{(\mathbf{m}^{\prime})}}%
\Delta\mathbf{x}_{\mathbf{mm}^{\prime}} \label{delx}%
\end{align}
Here, the summation $\sum_{{(\mathbf{m}^{\prime})}}$ stands for the summation
over the position indices of the four nearest neighbor beads of the bead
initially located at $\mathbf{m}$. Note that $\Delta\mathbf{x}_{\mathbf{m}}$
is nonlocal for the lattice index, and thus the equations of motions in the
above couple the dynamics of a bead with its nearest neighbors (see Sec.
\ref{s0} of Appendix for the details). The force $\mathbf{F}_{\mathbf{m}}$ is
the external force acting on the beads at $\mathbf{x}_{\mathbf{m}}$. This
force is set to zero except for the case in which we consider the boundary
force (as in the case of a creep test). These models describe a strongly
viscous case in which the inertial term is neglected and thus the elastic and
viscous stress are always balanced in the system when $\mathbf{F}_{\mathbf{m}%
}=0$.

When we stretch the two-dimensional network in the $y$ direction, we set
initially the $x$-component of $\Delta\mathbf{x}_{\mathbf{mm}^{\prime}}$ to
zero for simplicity. The $y$ component of the above equations of motion are,
respectively, given as%

\begin{align}
\gamma\frac{d}{dt}\Delta y_{\mathbf{m}}+k\Delta y_{\mathbf{m}}  &
=Y_{\mathbf{m}}\label{single}\\
\gamma\frac{d}{dt}y_{\mathbf{m}}+k\Delta y_{\mathbf{m}}  &  =Y_{\mathbf{m}}
\label{multi}%
\end{align}
where $\Delta y_{\mathbf{m}}$ and $Y_{\mathbf{m}}$ are the $y$ components of
$\Delta\mathbf{x}_{\mathbf{m}}$ and $\mathbf{F}_{\mathbf{m}}$, respectively.

The first case whose dynamics is governed by Eq. (\ref{single}) will be called
the model with a single relaxation time, whereas the second case in Eq.
(\ref{multi}) the model with multi relaxation times. This is because as
explained in Sec. \ref{s1} of Appendix, the former model possesses a single
relaxation time $\tau=\gamma/k\,$, whereas the latter has $N$ relaxation times
$\tau_{i}=\gamma/(k\lambda_{i})$ with $i=1,\ldots,N$. [As demonstrated in Sec.
\ref{s1} of Appendix, the maximum of $\tau_{i}$ scales with $N^{\nu}\gamma/k$
where $\nu$ is close to 2 and the minimum approach $(5/16)\gamma/k$ as $N$
increases.] The response to the creep test of the first model is equivalent to
that of Voigt model, whereas rheological properties of the second model are
similar to those of cross-linked polymers. Further details on rheological
aspects are discussed in Sec. \ref{s2} of Appendix, in which rheological
properties of the two models are examined.

For convenience, we introduce elastic modulus $E$ (of the unit Pa) and
viscosity \ $\eta$ (of the unit Pa$\cdot$s) through relations $E=k/d$ and
$\eta=\gamma/d$. The local strain for a given spring is given by the
elongation of the spring divided by the lattice spacing $d$. The global strain
for the network is defined as the elongation of the distance between the top
and bottom rows $\Delta L$ divided by the original length $L$ (see Fig.
\ref{Fig1}).

The numerical calculation for crack-propagation tests are performed as
follows. In the static test, we first prepare an equilibrium state for the
network with an initial (homogeneous) strain in the $y$ direction by giving
fixed positions at the top and bottom of the network; second, we introduce a
crack of length $a_{0}$ by removing springs from network located at
corresponding positions; third, we move each bead in the network on the basis
of the equation of motion, i.e., either of eq. (\ref{single}) or eq.
(\ref{multi}), under the condition that any spring in the network is removed
if the strain of the spring reaches a critical value $\varepsilon_{c}$.

In the dynamic test under constant-speed stretching at the velocity $U$, we
first introduce a crack of length $a_{0}$ in the network system in the
unstretched state (in which the length of each spring is equal to its natural
length) by removing springs; second, we start to move all the beads at the top
row upwards (i.e., in the positive $y$ direction) with the speed $U$, while
each bead moves on the basis of the equation of motion under the condition
that any spring in the network is removed if the strain of the spring reaches
a critical value $\varepsilon_{c}$.

In both cases, we only solve the $y\,$-component equation because the $x$ and
$y$ components are decoupled in Eqs. (\ref{single0}) and (\ref{multi0}). In
addition, we set $Y_{\mathbf{m}}=0$ because the boundary condition is given
not by force but by strain in both cases of the static and dynamic tests (when
we consider the creep test in Sec. \ref{s2} of Appendix, we deal with the case
of nonzero $Y_{\mathbf{m}}$).

In the model with multi relaxation times, the extension is transmitted from
the top to bottom of the system with time delay (note that the top row is in
tension while the bottom is fixed). Concerning this time delay, we confirmed
numerically the following property. In the case without crack, i.e., when
every column is stretched in the same way, if we introduce the difference
between the strains at adjacent (with respect to the $j$ index) rows by
$\Delta\varepsilon_{j}=\varepsilon_{j+1}-\varepsilon_{j}$ for the homogeneous
(with respect to the $i$ index) strain $\varepsilon_{j}=y_{i,j+1}-y_{i,j}$,
this quantity $\Delta\varepsilon_{j}$ satisfies the relation $\Delta
\varepsilon_{j}=\eta Uj/(NE)$. This implies that if the quantity $\eta U/(NE)$
is not small enough, it is possible that the strain of the spring at the top
$\varepsilon_{N-1}$ reaches $\varepsilon_{c}$ before the strain at the crack
tip does. In such a case, the crack does not propagate from the tip, but the
network starts to break near the top row. Analytical expressions for the model
concerning related properties, including its rheological functions, will be
discussed elsewhere.

\section{Results and Discussions}

The simulation is performed under the standard parameter set $(E,\eta
,\varepsilon_{c},L,d)=(100,80,0.32,10,1)$ unless specified (in appropriate
dimensionless units; e.g., the unit of $L$ is $d$). The width of the system
$W$ and the initial crack size $a_{0}$ are set to $1000d$ and $100d$ by
default in both of the simulation models. However, in the dynamic test, when
$U$ is relatively large, $W$ is made larger than the default value up to
$6000d$ for the global strain to reach $\varepsilon_{c}$ before the crack tip
reaches the opposite side edge of the sample (in order to observe the least
upper bound discussed below). In the following, we show plots of the crack
propagation velocity $V$ as a function of the energy release rate $G$, which
is given by $E\varepsilon^{2}L/2$ in the present linear case.

\subsection{Model with a single relaxation time}

In this section, we compare the results of crack-propagation tests under the
static and dynamic boundary conditions obtained from the model with a single
relaxation time. We announce here in advance that our numerical results
provided below support the following conclusion. If the condition%
\begin{align}
U  &  <U_{c}\label{eq1}\\
U_{c}  &  =Ed/\eta\label{eq1a}%
\end{align}
is well satisfied the static and dynamic tests give the same result. On the
contrary, when $U$ does not satisfy this condition, the plot of the
crack-propagation speed $V$ as a function of a given strain $\varepsilon$
obtained from the dynamic test tends to shift upwards as $U$ increases. This
may be understood as follows. Near the crack tip, the two different dynamics,
both tend to increase the strain $\varepsilon$ $(<\varepsilon_{c})$ of the
spring at the crack tip, compete with each other: one associated with the
relaxation characterized by the time scale $\eta/E$ and the other associated
with the strain rate set by the pulling speed $U$ characterized by the time
$d/U$. (Note that in the model with a single relaxation time the network is
homogeneously stretched if cracks are absent and the stretching motion is
characterized by the strain rate $U/d$.) As a result, as long as the
relaxation dynamics governs the increase of the strain at the crack tip, which
is equivalent to the condition for the time scales $\eta/E<d/U$, the results
of the dynamic test become equivalent to those of the static test. In fact,
this time-scale condition is identical to the condition given in Eq.
(\ref{eq1}).

\begin{figure}[h]
\includegraphics[width=\textwidth]{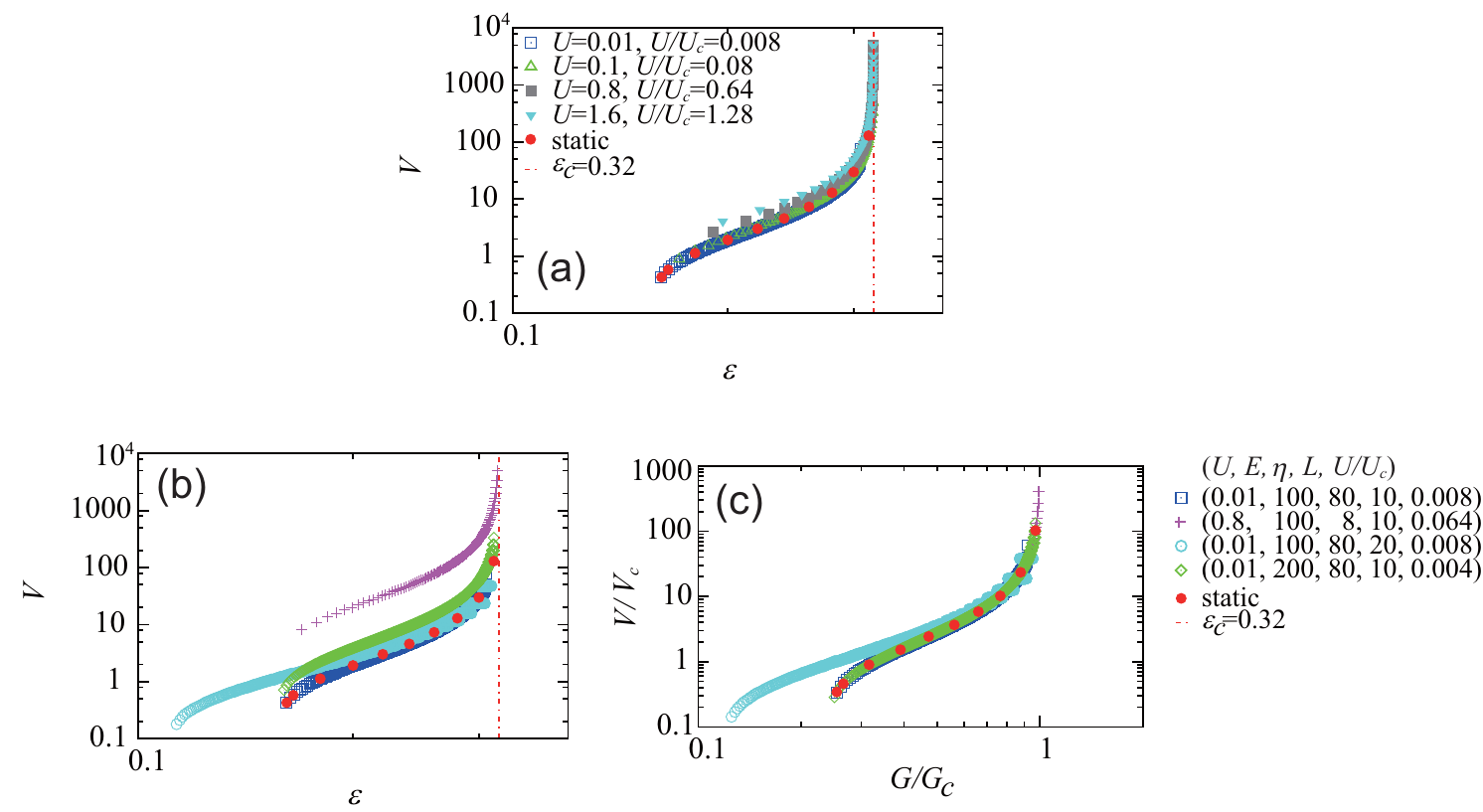}\caption{{}Results from the model
with a single relaxation time. (a) $V$ vs $\varepsilon$ for different pulling
speed $U$ for the fixed parameter set $(E,\eta,\varepsilon_{c}%
,L,d)=(100,80,0.32,10,1)$, from which $U_{c}$ is estimated as 1.25. (b) $V$ vs
$\varepsilon$ for different pulling speed $U$ for the different parameter sets
with the condition $U/U_{c}<1$ well satisfied, where $(\varepsilon
_{c},d)=(0.32,1)$. (c) The data shown in (b) on the renormalized axes
$V/V_{c}$ and $G/G_{c}$.}%
\label{Fig2}%
\end{figure}

Now, we confirm the above physical arguments by our numerical data obtained
from simulation. In Fig. \ref{Fig2} (a), the relation between the
crack-propagation speed $V$ and the global strain of the system $\varepsilon$
are given. The result obtained from the static test is compared with those
from the dynamic test performed under different stretching speed $U$. The data
are obtained for the same, standard parameter set $(E,\eta,\varepsilon
_{c},L,d)=(100,80,0.32,10,1)$. Thus, any differences in the results come from
differences in the boundary conditions. As shown in the plot, we confirm that
if the condition in Eq. (\ref{eq1}) is well satisfied the results from the
static and dynamic tests are the same, whereas slight upwards shifts are
observed with the increase in $U$ if Eq. (\ref{eq1}) is not well satisfied.
The reason of shifts in the upwards direction can physically be understood if
we remind that the shifts originate from the fact that pulling with the
velocity $U$ expedites faster stress concentration at the crack tip.

In Fig. \ref{Fig2} (b) and (c), we show that the above results is not special
to the standard parameter set and that the results obtained at a small
velocity $U$ which satisfies Eq. (\ref{eq1}) well possess the following
properties, which the results in the static test were confirmed to satisfy in
our previous study \cite{aoyanagi2017,aoyanagiNonlinear}. (A) $V/V_{c}$ is
given as a function of $G/G_{c}$, where%
\begin{align}
V_{c}  &  =Ed/\eta\label{eq3}\\
G_{c}  &  =E\varepsilon_{c}^{2}L/2 \label{eq4}%
\end{align}
This implies that the $V-\varepsilon$ curves collapse on to a single master
curve when plotted on the renormalized axes, $V/V_{c}$ and $G/G_{c}$. The
velocity scale $V_{c}$ is given by the smallest length scale $d$ divided by
the single time scale of the model $\eta/E$, i.e., $V_{c}$ is the smallest
velocity scale of the model. The quantity $G_{c}$ is the value of $G$
evaluated when $\varepsilon$ matches its critical value $\varepsilon_{c}$,
i.e., $G_{c}$ is the largest scale of $G$. (B) The greatest lower bound and
least upper bound for the $V/V_{c}-G/G_{c}$ relation are characterized by
$G_{\min}$ and $G_{\max}$ defined as%
\begin{align}
G_{\min}  &  =c_{1}G_{c}d/L=c_{1}E\varepsilon_{c}^{2}d/2\label{eq5}\\
G_{\max}  &  =G_{c} \label{eq6}%
\end{align}
with a universal numerical coefficient $c_{1}$ (see below). Here, for later
convenience, we define $\varepsilon_{0}$ by the following equation:%
\begin{equation}
G_{\min}=E\varepsilon_{0}^{2}L/2 \label{eq5b}%
\end{equation}
It is natural that $G_{\max}$ is given by $G_{c}$ considering that the spring
breaks when $\varepsilon=\varepsilon_{c}$. The bound $G_{\min}$ corresponds to
the static fracture energy discussed by Lake and Thomas \cite{LakeThomas1967},
and also corresponds to the critical state in which the maximum stress at the
crack tip coincides with the intrinsic failure $E\varepsilon_{c}$.

In Fig. \ref{Fig2} (b), we show the results obtained from various parameters
but with the stretching velocities $U$ all satisfy Eq. (\ref{eq1}) well.
Although the data in Fig. \ref{Fig2} (b) are scattered, when the same data are
replotted on the renormalized axes based on Eqs. (\ref{eq3}) and (\ref{eq4}),
they collapse onto a master curve: the dynamic test has properties (A) and
(B), as the static test does. The deviation of the data with $L=20$ in Fig.
\ref{Fig2} (c) is consistent with Eq. (\ref{eq5}), where $c_{1}$ is
approximately 2.3 for all the data shown in Fig. \ref{Fig2}. (The constant
$c_{1}$ seems weekly dependent on $N$ but takes the same value for the two
simulation models for a given $N$; in the previous studies
\cite{aoyanagi2017,aoyanagiNonlinear}, $c_{1}$ is approximately 2.4 in the two
simulation models for $N=200$, which is about ten times larger than present
values of $N$.)

\subsection{Model with multi relaxation times}

In this section, we compare the results of crack-propagation tests under the
static and dynamic boundary conditions obtained from the model with multi
relaxation times. We announce here in advance that our numerical results
provided below in Fig. \ref{Fig3} support the following properties. If the
following condition is well satisfied the static and dynamic tests give the
same result:%
\begin{align}
U  &  <U_{c,m}\label{eq7}\\
U_{c,m}  &  =U_{c}/N=(Ed/\eta)/N
\end{align}
When $U$ does not satisfy this condition in the dynamic test, the least upper
bound and greatest lower bound for the strain $\varepsilon_{\max}$ and
$\varepsilon_{\min}$, which are $\varepsilon_{c}$ and $\varepsilon_{0}$ [in
Eq. (\ref{eq5b})] in the static test, decreases and increases, respectively.
These two bounds are given by the following relations as shown in Fig.
\ref{Fig3} (d) below:
\begin{align}
\Delta_{\max}  &  =\frac{\varepsilon_{c}-\varepsilon_{\max}}{\varepsilon_{c}%
}=c_{2}\frac{U}{U_{c,m}}\label{eq8}\\
\Delta_{\min}  &  =\frac{\varepsilon_{\min}-\varepsilon_{0}}{\varepsilon_{0}%
}=c_{2}\frac{U}{U_{c,m}} \label{eq8b}%
\end{align}
with a universal constant: $c_{2}=0.9\pm0.003$. Note that the least upper
bound and the greatest lower bound in the dynamic test, $\varepsilon_{\max}$
and $\varepsilon_{\min}$, given respectively through Eqs. (\ref{eq8}) and
(\ref{eq8b}), approach the values $\varepsilon_{c}$ and $\varepsilon_{0}$ of
the static test when Eq. (\ref{eq7}) becomes well satisfied.

If we consider the two competing dynamics near the crack tip as in the case of
the model with multi relaxation times, the condition under which the results
of the dynamic test agrees with those of the static test is expected to be
given by the following: the longest relaxation time of the multi model, which
scales as $(\eta/E)N^{2}$ as shown in Sec. \ref{s1} of Appendix, is shorter
than the time scale of the strain rate set by the stretching velocity $U$,
which is in this case $L/U$. (Note that in the multi model the network is
inhomogeneously stretched even if cracks are absent and the stretching motion
is characterized not by the strain rate $U/d$ but by $U/L$.) This condition
for the time scales reduces to the condition given in Eq. (\ref{eq7}).

\begin{figure}[h]
\includegraphics[width=\textwidth]{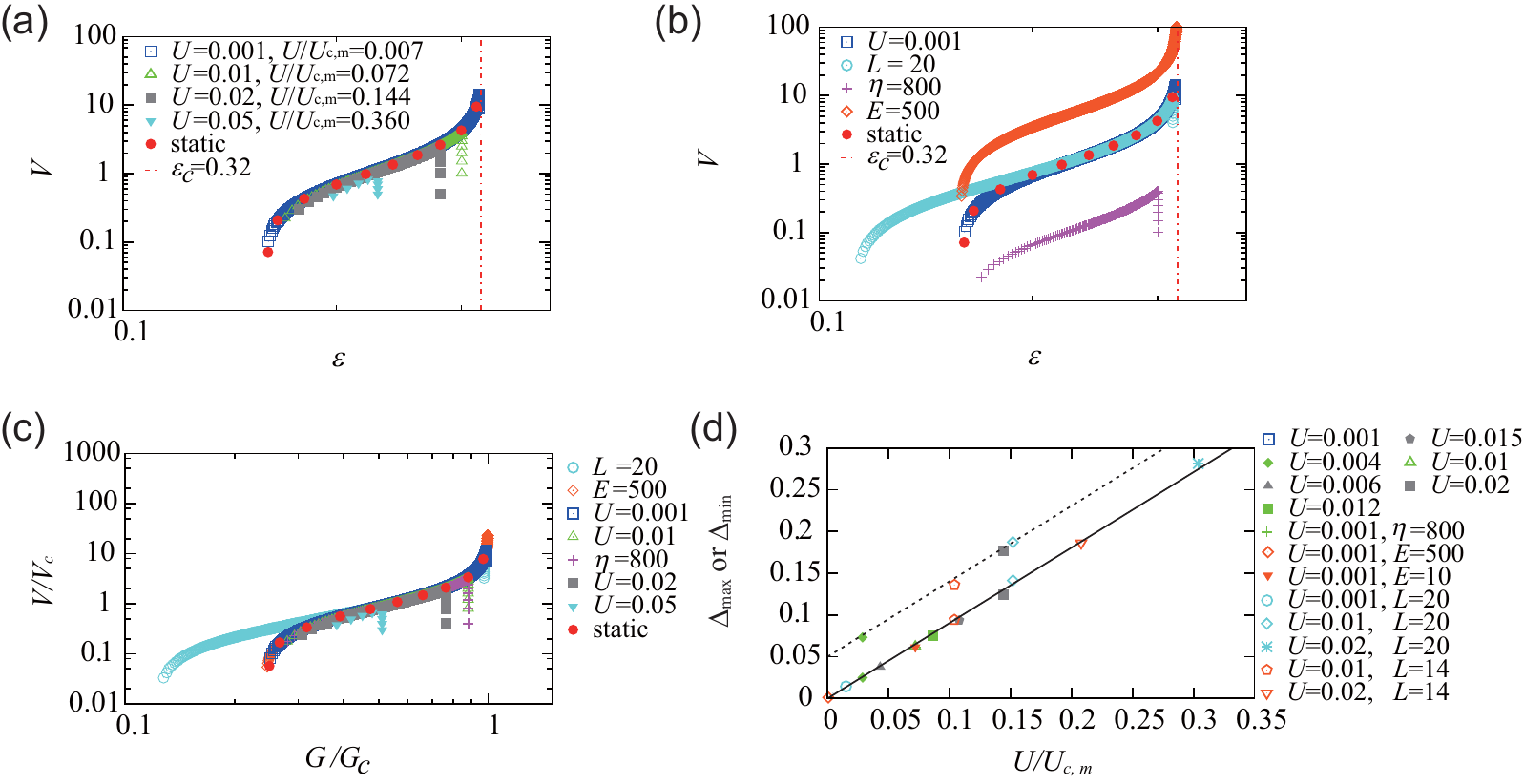}\caption{{}Results from the model
with multi relaxation times. (a) $V$ vs $\varepsilon$ for different pulling
speed $U$ for the fixed parameter set $(E,\eta,\varepsilon_{c}%
,L,d)=(100,80,0.32,10,1)$, from which we obtain $U_{c,m}=0.139$. (b) $V$ vs
$\varepsilon$ for the same pulling speed $U=0.001$ for different parameter
sets. In the legends, the only parameter changed in the standard set given in
the caption to (a) are given [except for the data labeled as $U=0.001$, which
is the same data used in (a)]. (c) The data shown in (b) on renormalized axes
$V/V_{c}$ and $G/G_{c}$. (d) Renormalized deviation of strains $\Delta_{\min}$
or $\Delta_{\max}$ defined in the text as a function of $U/U_{c,m}$. (For
clarity, $\Delta_{\min}$ is shifted upwards by 0.05; the dashed and solid
lines correspond to $\Delta_{\min}$ in Eq. (\ref{eq8b}) and $\Delta_{\max}$ in
Eq. (\ref{eq8}), respectively.) In the legends, the data for which only $U$ is
shown are performed for the standard parameter set given in the caption to
(a). The data for which $U$ and another parameter are given are performed for
the standard set but with the parameter shown in the legend (except $U$) is
replaced by the value.}%
\label{Fig3}%
\end{figure}

Now, we confirm the above physical arguments by our numerical data obtained
from simulation. In Fig. \ref{Fig3} (a), the relations between the
crack-propagation speed $V$ and the global strain of the system $\varepsilon$
are given for the static and dynamic tests. The data are obtained for the
same, standard parameter set $(E,\eta,\varepsilon_{c},L,d)=(100,80,0.32,10,1)$
as before so that differences reflect differences in the boundary conditions.
As shown in the plot, we confirm that as the condition in Eq. (\ref{eq7}) is
less satisfied the least upper-bound strain $\varepsilon_{\max}$ and the
greatest lower-bound strain $\varepsilon_{\min}$ for the $V-\varepsilon$
relation decreases and increases, respectively. (In the plot, $V$ looks
scattered at $\varepsilon=\varepsilon_{\max}$; in fact, the highest $V$ value
at $\varepsilon=\varepsilon_{\max}$, which is on the smooth curve suggested by
the $V$ values at smaller epsilons, is the model prediction, while the other
smaller $V$ values at $\varepsilon=\varepsilon_{\max}$ are added for guide for
the eyes to recognize the position of $\varepsilon=\varepsilon_{\max}$.) In
Fig. \ref{Fig3} (b) and (c), we show that the above results are not peculiar
to the standard parameter set, and that the results obtained even at a
velocity $U$ which does not satisfy Eq. (\ref{eq7}) possess properties (A) and
(B) but with $G_{\min}=E\varepsilon_{0}^{2}L/2$ and $G_{\max}=E\varepsilon
_{c}^{2}L/2$, replaced by $G_{\min,c}$ and $G_{\max,c}$, respectively, with%
\begin{align}
G_{\min,c}  &  =E\varepsilon_{\min}^{2}L/2\\
G_{\max,c}  &  =E\varepsilon_{\max}^{2}L/2\nonumber
\end{align}
Here, $\varepsilon_{\min}$ and $\varepsilon_{\max}$ satisfy Eqs. (\ref{eq8b})
and (\ref{eq8}), respectively. In Fig. \ref{Fig3} (d), the relations given in
Eqs. (\ref{eq8}) and (\ref{eq8b}) are directly confirmed. These relations are
quite natural as their simplest forms. This is because we generally expect
that the dimensionless quantities on the left-hand sides that measure the
deviations of the dynamic test from the static test, $\Delta_{\max}$ and
$\Delta_{\min}$, should scale with the positive power of the dimensionless
expression on the right-hand side $U/U_{c,m}$, which can naturally be
constructed from the condition in Eq. (\ref{eq7}). Note that the upwards shift
of the $V-\varepsilon$ plot is not observed in the case of the model with
multi relaxation times because the given stretching velocities $U$ are much
smaller than those in the model with a single relaxation time.

\subsection{Velocity jump in the static and dynamic tests}

By employing the mechanism of velocity jump elucidated in the previous
analytical theory \cite{sakumichi2017exactly}, we can reproduce the velocity
jump in our simple models in an \textit{ad hoc} manner. The theory predicts
that the glass transition that is very localized near the crack tip triggers
the jump. Accordingly, to reproduce the velocity jump, we have only to vitrify
the crack tip column by changing the elastic modulus of the springs in the
column from $E$, corresponding to the rubbery modulus, to the glassy modulus
$E_{G}$, if the strain exceeds the critical strain at the jump $\varepsilon
_{J}$.

On the basis of the physical arguments we have developed above concerning the
conditions for the static and dynamic tests to coincide with each other for
the two simulation models, we expect that the positions just before and after
the jump on the $V/V_{c}-G/G_{c}$ plot, characterized by the two points
$(V_{b}$, $G_{J})$ and $(V_{a}$, $G_{J})$, are the same in the static and
dynamic test if the condition in Eq. (\ref{eq1}) or Eq. (\ref{eq7}) is well satisfied.

\begin{figure}[h]
\includegraphics[width=\textwidth]{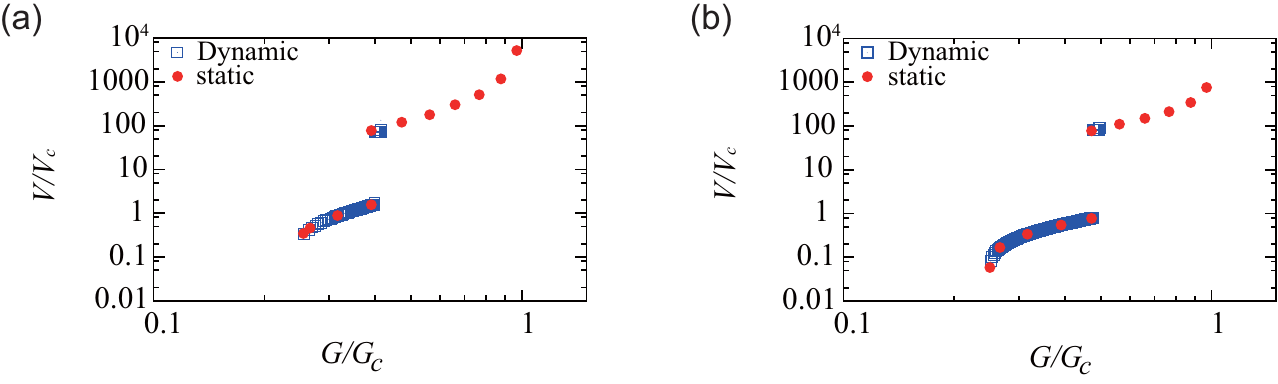}\caption{Comparison of the
relation between $V/V_{c}$ and $G$\thinspace/$G_{c}$ in the static and dynamic
tests obtained for the parameter set $(E,\eta,\varepsilon_{c}%
,L,d)=(100,80,0.32,10,1)$. (a) Model with a single relaxation time where
$U/U_{c}=0.008$. (b) Model with multi relaxation times where $U/U_{c,m}%
=0.007$.}%
\label{Fig4}%
\end{figure}

This expectation is confirmed in Fig. \ref{Fig4}, in which Eq. (\ref{eq1}) and
Eq. (\ref{eq7}) are well satisfied for the models with a single relaxation
time and multi relaxation times, respectively, in the dynamic test. As seen
the figure, if Eq. (\ref{eq1}) is well satisfied in the dynamic test for the
model with a single relaxation time, the fracture energy $G_{J}=E\varepsilon
_{J}^{2}L/2$ at the velocity jump, the velocities just before and after the
transition $V_{b}$ and $V_{a}$ ($V_{b}<V_{a}$) are the exactly the same in the
two tests (Here, we set $\varepsilon_{J}=0.2$ or $G_{J}/G_{c}=0.39$, and
$E_{G}=100E$). The same is true for the model with multi relaxation times, if
the condition given in Eq. (\ref{eq7}) is satisfied (Here, we set
$\varepsilon_{J}=0.22$ or $G_{J}/G_{c}=0.47$, and $E_{G}=100E$).

The range of the high velocity regime after the jump observed in the dynamic
test is very narrow. This comes from the practical limitation for the width
$W$ of the sample. If the width were long enough we would have the same range
of the high velocity regime in the dynamic test, if Eq. (\ref{eq1}) and Eq.
(\ref{eq7}) are well satisfied. After the jump, velocity increases
significantly. Because of this, the crack tip soon reaches the opposite side
edge of the sample in practice. The range of the low velocity regime before
the jump could also become narrower (in the model with multi relaxation times)
if $U$ is not small enough because the greatest lower bound increases
according to Eq. (\ref{eq8b}).

In our previous work (see Fig. 5 and 6 of \cite{sakumichi2017exactly}), we
showed that only the very vicinity of the crack tip is vitrified during
stationary crack propagation at velocities above the velocity transition. This
implies that the speed of vitrification is faster than any relaxation dynamics
in the system, and that the strong tension transmitted from the boundary of
vitrified region contributes to accelerate the crack propagation speed by
expediting stress concentration around the crack tip to reach the critical
strain $\varepsilon_{c}$. In this sense, our artificial manner of
vitrification physically mimics the mechanism of accelerated crack propagation
due to vitrification. In addition, it should be noted that in the static test
the simulations at strains below and above vitrification are performed
independently. Even in the dynamic test, the situation is essentially the same
because we focus on a slow pulling-velocity regime, in which relaxation
dynamics relevant for crack propagation is much faster than the time scale
that characterizes the pulling speed. Note also that the energy release rate
is defined based on homogeneous elastic energy developed well away from the
crack tip and thus well defined in both cases of below and above transition.

\section{Conclusion}

In this study, we showed that the relation between the crack-propagation
velocity $V$ and the strain $\varepsilon$ or the fracture energy (i.e., energy
release rate) $G$ obtained from the static and dynamic tests are the same if
the stretching velocity $U$ is smaller than a critical value, which we
generally call $U_{CR}$ for later convenience. This is true even if the
velocity jump exists. However, when the velocity jump is present, it is
practically difficult to observe the full range of the high velocity regime
after the jump because of the finite width of actual samples. (The small
velocity regime also tends to be narrower if $U$ is not small enough.)

The present study provided the critical values $U_{CR}$ for the two simulation
models [see Eqs. (\ref{eq1}) and (\ref{eq7})] and their physical
interpretation. On the basis of the interpretation, the critical value is
universally given by the condition that the longest relaxation time of the
sample is shorter than the time scale of stretching $L/U$ (because any
practical system is the system with multi relaxation times). This condition
could be hard to satisfy in practice for a system with long chains. However,
this condition is relaxed if our purpose is limited to the detection of the
velocity jump. This is because the jump positions just before and after jump,
$(V_{b}$, $G_{J})$ and $(V_{a}$, $G_{J})$ with $G_{J}=E\varepsilon_{J}^{2}%
L/2$, are the same only if the strain at the jump $\varepsilon_{J}$ is in the
range of the dynamic test, $\varepsilon_{\min}<\varepsilon<\varepsilon_{\max}$
, which is narrower than the range of the static test, $\varepsilon
_{0}<\varepsilon<\varepsilon_{c}$.

The dynamic test is promising for the study of the velocity jump including non
rubbery polymers because of the timesaving and cost-effective features and
high sensitivity for detecting the velocity jump. Considering that the
velocity jump has been utilized effectively for developing tough elastomers in
the industry, the results of the present study are useful for developing tough
polymers in general. This is especially because the previous study
\cite{sakumichi2017exactly} suggests theoretically that the velocity jump is
expected be observed universally for variety of materials and the previous
studies \cite{TomizawaJump2} and \cite{Takei2018} demonstrate experimentally
that the dynamic test is more sensitive for detecting the velocity jump.

\begin{acknowledgments}
This work was partly supported by ImPACT Program of Council for Science,
Technology and Innovation (Cabinet Office, Government of Japan).
\end{acknowledgments}


\newpage

\appendix

\section*{Appendix}

\subsection{Details of the simulation models}

\label{s0}

In the lattice network, the four nearest neighbor cites of the cite
$\mathbf{m}=(i,j)$ are specified by the indices $\mathbf{m}^{\prime}$ $=$
$(i-1,j),$ $(i,j+1),$ $(i+1,j),$ and $(i,j-1)$, which are called
$\mathbf{m}_{s}$ with $s=1,$ 2, 3, and 4, respectively. The elongation vector
$\Delta\mathbf{x}_{\mathbf{mm}^{\prime}}$ are defined as $\Delta
\mathbf{x}_{\mathbf{mm}_{s}}=$ $\mathbf{x}_{\mathbf{m}}-\mathbf{x}%
_{\mathbf{m}_{s}}-\mathbf{d}_{s}$ with $\mathbf{d}_{1}=(d,0)=-\mathbf{d}_{3}$
and $\mathbf{d}_{4}=(0,d)=-\mathbf{d}_{2}$.

As mentioned in the text, in both of the two models in Eqs. (\ref{single0})
and (\ref{multi0}), the dynamics of a bead is coupled with its nearest
neighbors. For example, in the first case, the simulation algorithm is as
follows. (i) We renew "$\Delta\mathbf{x}_{\mathbf{m}}$ at the $i$-th step" on
the basis of Eq. (\ref{single0}) to obtain "the renewed $\Delta\mathbf{x}%
_{\mathbf{m}}$." (ii) "The renewed $\Delta\mathbf{x}_{\mathbf{m}}$ is
transformed back to have "$\mathbf{x}_{\mathbf{m}}$ at the $(i+1)$-th step."
(iii) From this "$\mathbf{x}_{\mathbf{m}}$ at the $(i+1)$-th step", we
calculate "$\Delta\mathbf{x}_{\mathbf{m}}$ at the $(i+1)$-th step" using Eq.
(\ref{delx}). (iv) The "$\Delta\mathbf{x}_{\mathbf{m}}$ at the $(i+1)$-th
step" thus obtained is used for Eq. (\ref{single0}) to proceed to the
$(i+2)$-th step. Note here the following: (A) Eq. (\ref{delx}) implies that
$\Delta\mathbf{x}_{\mathbf{m}}$ is expressed as a linear combination of
$\mathbf{x}_{\mathbf{m}}$ and thus are connected by a matrix. (B) "The renewed
$\Delta\mathbf{x}_{\mathbf{m}}$" and "$\Delta\mathbf{x}_{\mathbf{m}}$ at the
$(i+1)$-th step" in the above are generally different because of the coupling
between $\mathbf{x}_{\mathbf{m}}$.

\subsection{Distribution of Relaxation times in the simulation models}

\label{s1}

To gain physical pictures of the models governed by Eqs. (\ref{single}) and
(\ref{multi}), we consider a creep test: we give a fixed stress $\sigma_{0}$
suddenly at time $t=0$ to observe the time development of the strain
$\varepsilon(t)$ after $t=0$. In such a case, each column (in the $y$
direction) behaves in the same way as its neighbors and the shear force does
not act at all: the problem reduces to one dimensional. For example, instead
of Eq. (\ref{single}), we have only to consider the equation%
\begin{align}
\gamma\frac{d}{dt}\Delta y_{\mathbf{m}}+k\Delta y_{\mathbf{m}}  &
=Y_{\mathbf{m}}\label{A1-1}\\
\Delta y_{\mathbf{m}}  &  =\Delta y_{\mathbf{mm}_{2}}+\Delta y_{\mathbf{mm}%
_{4}} \label{A1-1b}%
\end{align}
where $Y_{\mathbf{m}}=\sigma_{0}d^{2}$ for the beads at the top boundary and
$Y_{\mathbf{m}}=0$ for the remaining bead. If we seek the elementary solution
for Eq. (\ref{single}) of the form $\Delta y_{\mathbf{m}}{=y}_{0}e^{-t/\tau}$,
we find a single relaxation time $\tau=\gamma/k$. (The explicit solution for
the creep test is given below.)

\begin{figure}[h]
\includegraphics[width=\textwidth]{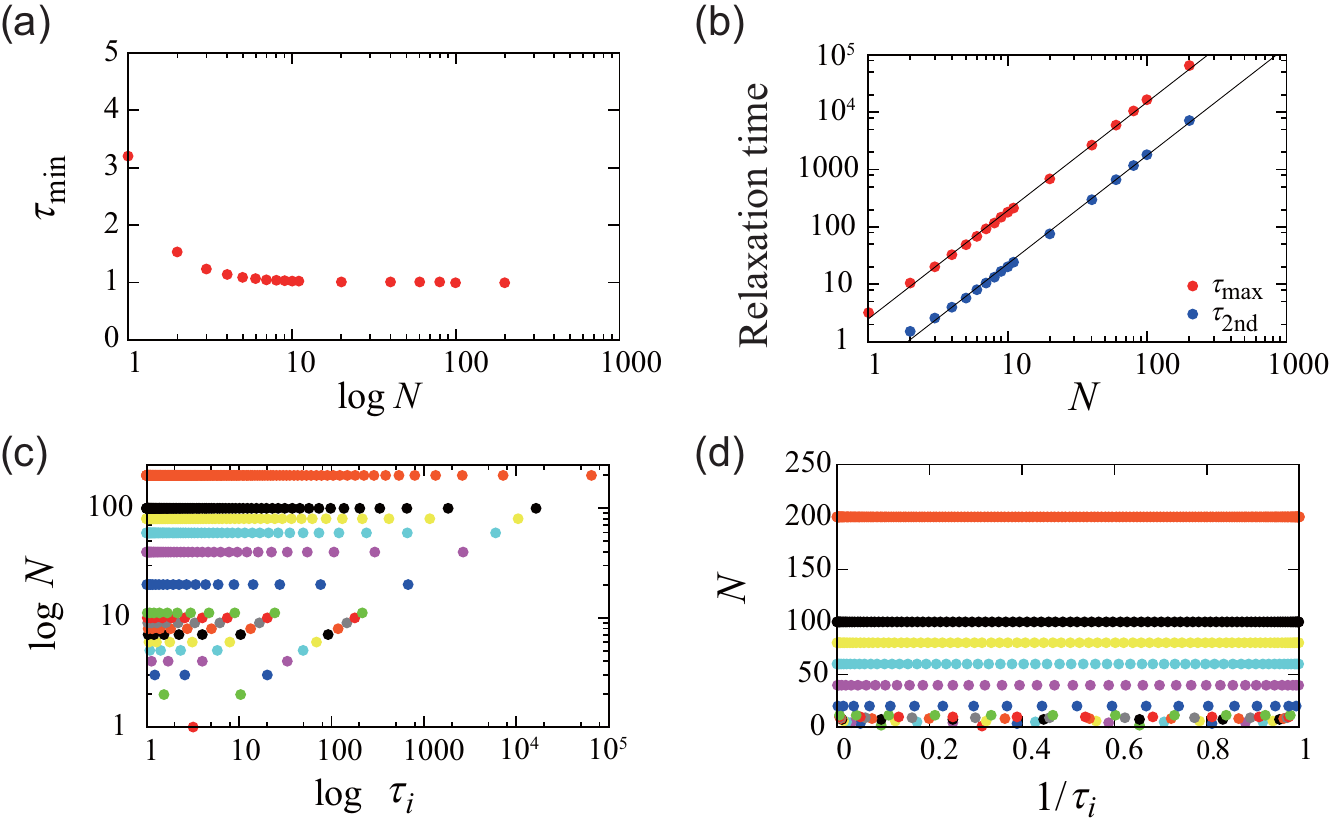}\caption{{}(a) The smallest
relaxation time $\tau_{\min}$ vs $N$. (b) The largest and second largest
relaxation times vs $N$. (c) Distribution of the relaxation time $\tau_{i}$ at
different $N$, where each dot represents one mode. (d) Distribution of the
inverse relaxation time $1/\tau_{i}$ at different different $N$. In the plots,
the unit of time in the plots are set to $\tau_{0}=(5/16)\gamma/k$ with
$\gamma/k=\eta/E$.}%
\label{FigA1}%
\end{figure}

On the contrary, in Eq. (\ref{multi}), we find $N$ mutually-dependent
equations
\begin{equation}
\gamma\frac{d}{dt}y_{\mathbf{mm}}+k\Delta y_{\mathbf{m}}=Y_{\mathbf{m}}
\label{A1-2}%
\end{equation}
where $\Delta y_{\mathbf{m}}$ and $Y_{\mathbf{m}}$ are given as before. (The
case of $N=1$ is special and is, in this specific case of the creep test,
identified with the model with a single relaxation time for convenience.) This
coupled equation set can be represented by a matrix formulation and the
relaxation times can be obtained by solving an eigen-value problem of the
matrix. By solving the eigen-value problem numerically, we obtained the
distribution of relaxation times and examined explicitly a few selected
characteristic relaxation times, the results of which are characterized in
Fig. \ref{FigA1}. The unit of time in the plots are set to $\tau
_{0}=(5/16)\gamma/k$, where the smallest relaxation time $\tau_{\min}$
approaches this value as $N$ increases, as shown in (a). The largest and
second largest relaxation times $\tau_{\max}$ and $\tau_{2nd}$ seem to scale
as $N^{\nu}$ with the exponent $\nu$ close to $2$ as shown in (b). The lines
in (b) represent the relations $\tau_{\max}/\tau_{0}=(2.49\pm0.057)\times
N^{1.89\pm0.0006}$ and $\tau_{2nd}/\tau_{0}=(0.189\pm0.008)\times
N^{1.89\pm0.0007}$, while we expect the exponent $\nu$ approaches 2 in the
large $N$ limit on the basis of the physical interpretation we provided for
Eq. (\ref{eq7}).

\subsection{Rheological properties of the simulation models}

\label{s2}

In this section, we demonstrate rheological properties of the two simulation
models. For convenience, we first review and define rheological functions. The
creep test is defined as follows: we give a fixed stress $\sigma_{0}$ suddenly
at time $t=0$ to observe the time development of the strain $\varepsilon(t)$
after $t=0$. From $\varepsilon(t)$ thus obtained, the extensional creep
compliance $D(t)$ is given as
\begin{equation}
\varepsilon(t)=D(t)\sigma_{0} \label{et}%
\end{equation}
The complex compliance $D^{\ast}(\omega)$ is introduced by the following
equation:
\begin{equation}
D^{\ast}(\omega)=i\omega\int_{0}^{\infty}D(t)e^{-i\omega t}dt \label{do}%
\end{equation}
This quantity and the complex modulus $E(\omega)$ satisfies the following
equation \cite{ferry1980viscoelastic}:
\begin{equation}
E^{\ast}(\omega)=\frac{1}{D^{\ast}(\omega)}=E^{\prime}(\omega)-iE^{\prime
\prime}(\omega) \label{eo}%
\end{equation}
By using Eqs. (\ref{et}) to (\ref{eo}), we can obtain rheological functions
$D(t)$, $E^{\prime}(\omega)$, and $E^{\prime\prime}(\omega)$ from the function
$\varepsilon(t)$ obtained from the creep test.

\subsubsection{Model with a single relaxation time}

On the basis of the equation of motion for the creep test in this model, given
in Eq. (\ref{A1-1}), we obtain $\varepsilon(t)=(\sigma_{0}/E)(1-e^{-t/\tau})$
with $\tau=\gamma/k=\eta/E$. Thus, from Eq. (\ref{et}), the creep compliance
is given by%

\begin{equation}
D(t)=\frac{1}{E}\left(  1-e^{-\frac{E}{\eta}t}\right)  \label{dts}%
\end{equation}

From Eq. (\ref{do}), we obtain $D(\omega)$ as
\begin{equation}
D(\omega)=\frac{\omega}{E\omega+i\eta\omega^{2}} \label{domega}%
\end{equation}
by assuming that $\omega$ has an infinitely-small negative imaginary part for
the integral to converge. From this expression, we obtain $E(\omega)$ from Eq.
(\ref{eo}):%

\begin{equation}
E(\omega)=\frac{E\omega+i\eta\omega^{2}}{\omega}%
\end{equation}
From this, we obtain%
\begin{equation}
E^{\prime}(\omega)=E \label{e1}%
\end{equation}%
\begin{equation}
E^{\prime\prime}(\omega)=\eta\omega\label{e2}%
\end{equation}

\subsubsection{Model with multi relaxation times}

For this model, we first obtain the function $\varepsilon(t)$ numerically for
the parameter set $(\eta,E)=(80,100)$, and fit the function with an analytical
expression in the following form with regarding $\tau_{n}$ as fitting
parameters and selecting $a$ for a given stress for the creep test:%
\begin{equation}
a\sum_{n}(1-e^{-t/\tau_{n}})
\end{equation}
On the basis of the analytical expression, we demonstrate plots of $D(\omega
)$, $E^{\prime}(\omega)$, and $E^{\prime\prime}(\omega)$ in Figs. \ref{FigA2}
and \ref{FigA3} below. For $N\leq12$, we used a sum of $N$ functions for the
fitting for simplicity ($a$ is set to 0.01 for a given stress $\sigma_{0}=20$,
corresponding to a saturation value of the strain at long times, $\sigma
_{0}/E=$ $0.2$). However, for $N>12$, we used a sum of only 12 functions for
convenience: our results below is subject to errors to a certain degree. More
complete examination will be discussed elsewhere.

\begin{figure}[h]
\includegraphics[width=\textwidth]{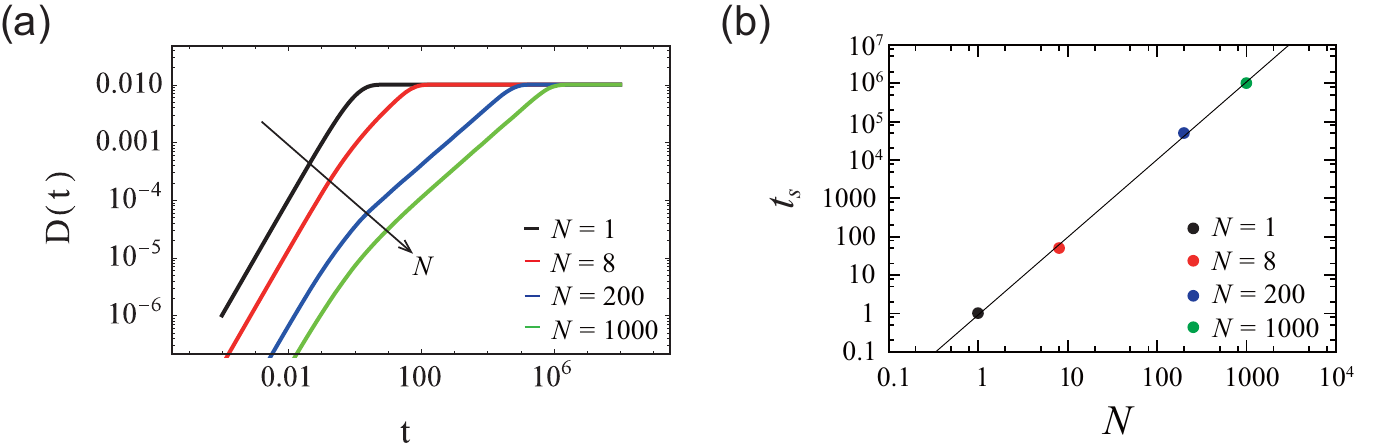}\caption{(a) Creep compliance
$D(t)$ for different $N$. (b) The saturation time $t_{S}$ as a function $N$.
In the plots, the unit of time is $\eta/E$ with $\eta=80$ and $E=100$.}%
\label{FigA2}%
\end{figure}

In Fig. \ref{FigA2}, we show the creep compliance $D(t)$ numerically obtained
based on Eq. (\ref{A1-2}). This function approaches the saturation value
$1/E=1/100$ at the saturation time $t_{S}$. This time is defined as the time
at the intersection of two extrapolations lines, one from the saturated
plateau region and the other from the region of straight line with positive
slope next to the plateau region. The time $t_{S}$ is given as a function of
$N$ in (b), in which the data are fitted by $t_{S}=c_{S}N^{\nu_{S}}$ with
$c_{S}=2.6\pm0.01$ and $\nu_{S}=1.86\pm0.00065$. Comparing this numerical
fitting with the one we obtained for $\tau_{\max}$ given in the previous
section, $t_{S}$ can physically be identified with $\tau_{\max}$, which
implies that the rheological functions are characterized by the longest
relaxation time $\tau_{\max}$.

\begin{figure}[h]
\includegraphics[width=\textwidth]{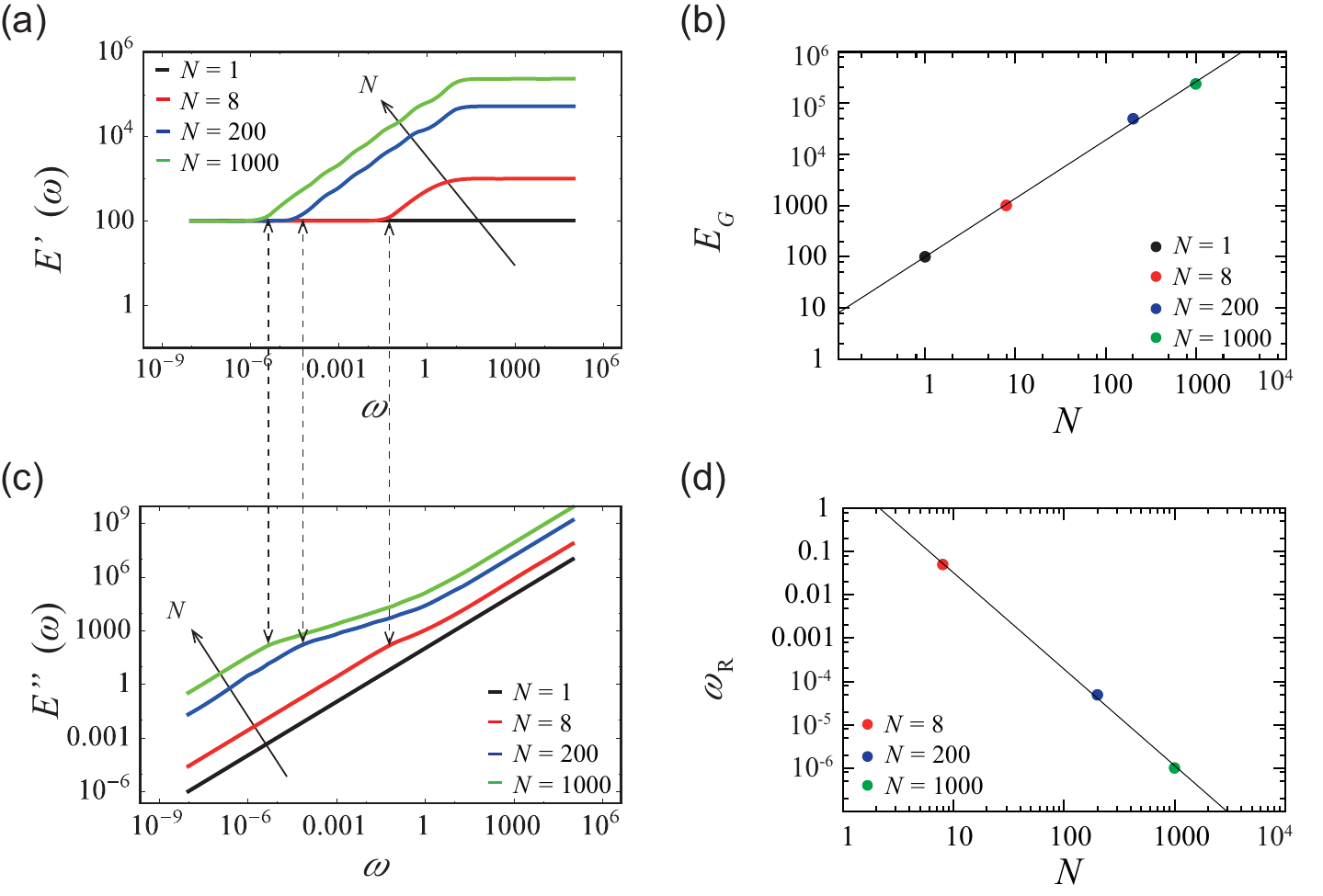}\caption{{}{}(a) $E^{\prime
}(\omega)$ for different $N$. (b) Glass modulus $E_{G}$ ($E^{\prime}(\omega)$
for large $\omega$) as a function of $N$. (c) $E^{\prime\prime}(\omega)$ for
different $N$. (d) Rubber frequency $\omega_{R}$ ($\omega$ at which the
initial rubbery plateau terminates) as a function of $N$. In the plots, the
unit of time is $\eta/E$ with $\eta=80$ and $E=100$, as before.}%
\label{FigA3}%
\end{figure}

In Fig. \ref{FigA3} (a), we show numerically obtained complex modulus
$E(\omega)$ for the model with multi relaxation times at different $N$. In
(a), we confirm that our model possesses the rubbery plateau $(E=100)$ on the
low frequency side and the glassy plateau on the high frequency side. The
glassy modulus $E_{G}$ increases with $N$, which is quantified in (b). The
data is fitted by $E_{G}=c_{G}N^{\nu_{G}}$ with $c_{G}=99.62\pm0.017$ and
$\nu_{G}=1.14\pm0.008$. The initial rubbery plateau is terminated at
$\omega=\omega_{R}$, at which $E^{\prime\prime}(\omega)$ starts to deviate
from the initial straight line as seen in (c). The rubbery frequency
$\omega_{R}$ decreases with $N$, which is quantified in (d). The data is
fitted by $\omega_{R}=c_{R}N^{\nu_{R}}$ with $c_{R}=5.48\pm0.032$ and $\nu
_{R}=-2.23\pm0.0036$. The characteristic time $1/\omega_{R}$ may be identified
with $\tau_{\max}$, which again suggests that the rheological functions are
characterized by the longest relaxation time $\tau_{\max}$.

\begin{figure}[h]
\includegraphics[width=\textwidth]{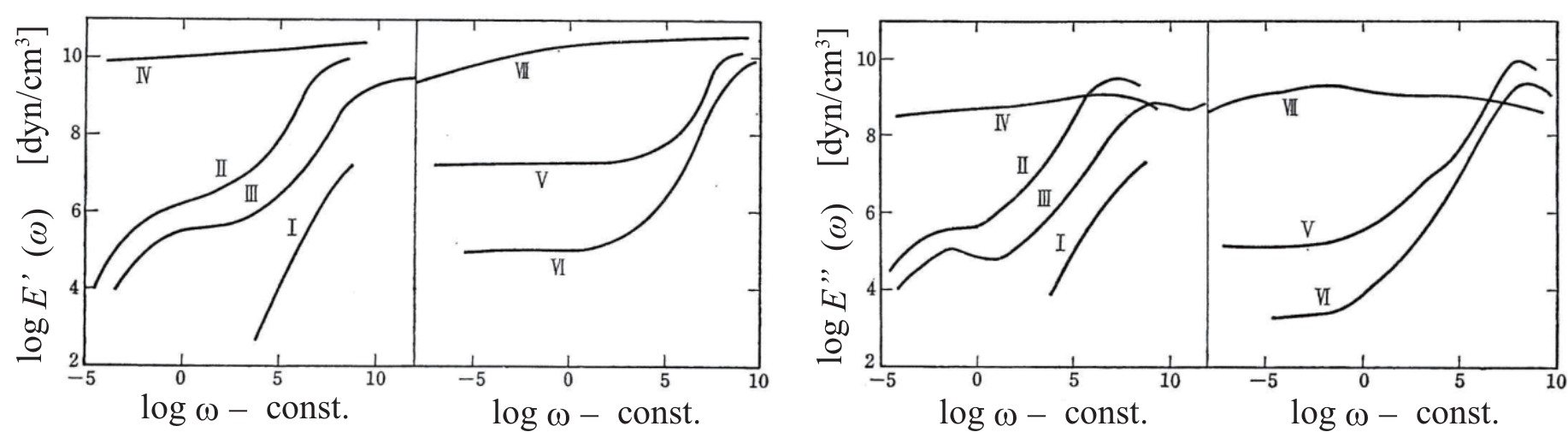}\caption{{}Typical plots of
$E^{\prime}(\omega)$ and $E^{\prime\prime}(\omega)$ obtained from polymers
introduced in the classic text \cite{ferry1980viscoelastic}.}%
\label{FigA4}%
\end{figure}

For completeness, we show in Fig. \ref{FigA4} typical rheology data for
various polymers. By comparing these plots and plots in Fig. \ref{FigA3}(a)
and (c), we see that our model with multi relaxation times is capable of
describing essential features of polymer rheologies.
\end{document}